\documentclass[11pt]{JHEP}
\usepackage{amssymb} 
\usepackage{amscd}

\def \be {\begin{equation}}
\def \ee {\end{equation}}
\def \bea {\begin{eqnarray}}
\def \eea {\end{eqnarray}}
\def \ba {\begin{array}}
\def \ea {\end{array}}
\def \pp {{\Bbb P}}

\def \cc {{\Bbb C}}
\def \oo {{\cal O}}
\def \qq {{\cal Q}}
\def \CY {Calabi-Yau }
\def \rrr{{\cal R}}
\newcommand {\mat}[1]{\mathop{\rm #1}\nolimits}

\newcommand{\ch}{\mathop{\rm ch}\nolimits}
\newcommand{\Hom}{\mathop{\rm Hom}\nolimits}
\newcommand{\Ext}{\mathop{\rm Ext}\nolimits}

\preprint{SISSA 100/2000/FM\\\tt hep-th/0010217}

\title{D-branes on Calabi-Yau manifolds and helices}

\author{A.~Tomasiello\\
International School for Advanced Studies (SISSA/ISAS)\\

Via Beirut 2--4, 34014 Trieste, Italy\\
E-mail: \email{tomasiel@sissa.it} }

\abstract{We investigate 
further on the correspondence between branes on a \CY in the
large volume limit and in the orbifold limit. We conjecture a new procedure
which improves computationally the McKay correspondence and prove it in a non
trivial example. We point out the relevance of helices and try to draw some
general conclusions about Beilinson theorem and McKay correspondence.}


\begin{document}

\section{Introduction}
The problem of describing D-branes in \CY manifolds is clearly central in
string theory, as a part of the general effort to understand its
non-perturbative dynamics. 
In the large volume limit the condition on the cycles they can wrap is known
\cite{Becker:1995kb} (for a review see \cite{Brunner:2000jq}).
So-called A branes are 3 dimensional (throughout this paper we will refer to
the dimension of the branes as if they were not extended in the uncompactified
dimensions) and wrap special Lagrangian submanifolds; B branes, on which we
will focus, are
even-dimensional and wrap holomorphic submanifolds. Of course this geometrical
description is deemed to become more complicated as soon as one tries to
travel in the interior of the moduli space; on the other hand, near the
orbifold point it is possible to apply the known methods to analyze branes on
an orbifold \cite{Douglas:1996sw}.\\
The generalization of these pictures to arbitrary points in the moduli space
turns out to be based on the language of categories and homological
algebra. Although still much is lacking to a complete understanding, it is
remarkable that some non--trivial results have been beginning to emerge
recently. 
As an example, a dictionary of branes between the large volume
and the orbifold points has been set up, based on generalized McKay
correspondence \cite{Diaconescu:2000ec,Diaconescu:2000dt}
and Beilinson theorem \cite{Diaconescu:2000ec,Douglas:2000qw}. 

In this paper we will investigate further in this direction. 
Motivated by strikingly simple results in \cite{Diaconescu:2000ec}, 
we will state and prove
in an example two conjectures: the first improves computationally the method,
while the value of the second is mainly theoretical, in that it makes one
discover 
the relevance of the concepts of helix and of mutation; these turn out to be
related in an interesting way to a Gram-Schmidt orthogonalization procedure
(actually this is already implicit in the work of Bondal \cite{semRu}).
We will note several
nice mathematical features and coincidences in the picture we will develop, and
try to draw a general lesson from it. 

We begin in section \ref{shqui} by reviewing needed physical and mathematical 
background. In section \ref{multiple} and \ref{GS} we state our two main
conjectures, and proceed to check them in explicit examples in section
\ref{expl}. There are moreover
interesting mathematical consequences that can be explored; one of this is the
comparison between McKay and Beilinson quivers, whose study we begin 
in section \ref{mkb}, for the case of projective spaces.
Finally, section \ref{disc} is an attempt to justify a posteriori the
success of our conjectures; we argue that the reason behind them
should be more a physical than a mathematical one.
\smallskip

\acknowledgments {This paper would not have seen the light without the
guide of Prof.~Michael R.~Douglas, who introduced me to these matters and
supported me throughout this work.
I also wish to thank IHES for its kind hospitality during
the initial phases of this project.}

\section{Sheaves and quivers}
\label{shqui}
As usual, we start from the linear sigma model approach \cite{Witten:1993yc}. 
Since in this paper we will only work with threefolds of codimension 1, 
this model has 5+1 chiral superfields $(Z^i, P)$, $i=1\ldots 5$ with 
charges $(w_i, K\equiv - \sum w_i)$ with respect to
the gauge group $U(1)$; the $D$-term and superpotential read
\be
D= \sum w_i |Z^i|^2 - K |P|^2 - \rho, \quad W=P \sum (Z^i)^{k_i + 2}.
\label{Dterm}
\ee
In the 
$\rho > 0$ phase, the moduli space of vacua reduces to a symplectic quotient 
easily recognizable as $\pp^{w_1,\ldots,w_5}$, and the $F$-flatness condition
becomes the equation which cuts a \CY. In the other phase, all $Z^i$ have
instead zero expectation value. In the extreme limits $\rho \to \pm \infty$, 
after integrating out the massive fields, both
theories flow to superconformal theories: the first one confines to a sigma
model on the \CY, the second one becomes a Landau-Ginzburg theory; actually an
orbifold thereof, due to an unbroken discrete relic of the $U(1)$ gauge group,
which acts on $Z^i$ as $\omega^{w_i}$, where $\omega= e^{2\pi i/K}$.

In both these theories we know how to describe even branes; 
in the geometric limit, they are holomorphic subvarieties with a holomorphic
(and stable) bundle on them; in the orbifold limit, we can apply the quiver
description  \cite{Douglas:1996sw}, and associate them to representations of a
quiver, with relations given by $F$-flatness.

On the other hand, the physical picture we have been reviewing is nothing but
a resolution of singularities. Indeed, 
the symplectic quotient ($D$-flatness plus
gauge invariance) of the first phase
is the exceptional locus of the resolution of the orbifold 
singularity of the second phase. \\
Given this simple observation, it is natural to think \cite{Diaconescu:2000ec}
that the correspondence
between branes in the two theories is given by McKay correspondence 
\cite{Reid, INa}. \\
Really, one of the points of this paper will be that this idea is naturally
connected with Beilinson
theorem \cite{beil}, which already turned out to be 
relevant in particular cases \cite{Douglas:2000qw}. For this reason we are now
going to briefly review both, of course with an eye on our applications.

\subsection{McKay correspondence}
What is classically called McKay correspondence is the correspondence between 
irreducible representations of a finite subgroup $\Gamma \subset SL_2(\cc)$ 
and 
irreducible components of the exceptional locus of the resolution of $\cc^2
/\Gamma$. For higher dimensions, the situation is not yet well understood; we
will mainly refer, anyway, to the concrete framework of \cite{INa}.

Let thus $\Gamma \subset SL_n(\cc)$. This inclusion means an action of $\Gamma$
on $\cc^n$, and this gives in turn a singularity $\cc^n/\Gamma$. Consider 
the irreducible representations $\rho_k$ of
$\Gamma$, and the $n$-dimensional representation $Q$ given by  
$\Gamma \subset SL_n(\cc)$. We can define a quiver whose dots represent
$\rho_k$, whose arrows are given by numbers $a_{ij}^{(1)}$ given by 
$Q \otimes \rho_j = \oplus \  a_{ij}^{(1)}\rho_i$ and with relations
which, in the case in which $\Gamma$ is abelian and cyclic, can be written  
$X_{a,a+w_i}^i X_{a+w_i,a+w_i+w_j}^j = X_{a,a+w_j}^j X_{a+w_j,a+w_i+w_j}^i$ 
(which, in a common
notation, can be written as $[X^i, X^j]=0$).\\
Let us now consider 
the space of representations of this quiver modulo 
its automorphisms,
\be
M\equiv \frac{\{ X \in (Q\otimes \mat{End} \rho)^\Gamma | [X^i, X^j]=0\}}
{\mat{GL}_\Gamma(\rho)}, \quad 
\mat{GL}_\Gamma(\rho)\equiv (\mat{End}\rho)^\Gamma
\label{resol}
\ee
with $\rho$ the regular representation 
(really this quotient should be defined more carefully as a GIT quotient
\cite{SI}). This turns out to be a
resolution of $\cc^n/\Gamma$, 
generalizing the 2-dimensional result of Kronheimer \cite{Kro}.\\
In a special case of this construction (like the resolutions obtained as
Hilbert schemes of points \cite{INa}), we can go further. Define $P$ as the
numerator of (\ref{resol}), and view it as a principal fibration $P\to M$;
then we can define the bundle associated to the regular fibration
$\rrr \equiv P \times_{\mat{GL}_\Gamma(\rho)}\rho$, and the ones associated 
to the irreps
$\rho_i$, $\rrr_i$; 
both are called 
tautological bundles.\\
Multiplication by the coordinates defines a complex
\be
\rrr \to Q \otimes \rrr \to \Lambda^2 Q \otimes \rrr \to \ldots 
\Lambda^n Q \otimes \rrr \cong \rrr,
\ee
which can be decomposed as
\be
\rrr_i \to \oplus a^{(1)}_{ji}\rrr_j \to 
\oplus a^{(2)}_{ji}\rrr_j \to \oplus a^{(3)}_{ji}\rrr_j \to \ldots \rrr_i
\ee
where 
$ \Lambda^k Q \otimes \rho_i = \oplus \  a_{ji}^{(k)}\rho_j$. What is
important for us is that this complex, that we call $S_i$, 
is exact outside the exceptional locus, and thus defines an element of its
K-theory; and that these $S_i$ are dual to the $\rrr_i$ in a sense compatible
with our definition of duality below.

Thus, from this point of view we have what we wanted: a map 
which associates to each irrep of $\Gamma$ $\rho_i$ a K-theory class on the
exceptional divisor (and hence, by restriction, on the \CY).

\subsection{Beilinson theorem}
\label{Bth}
This construction \cite{beil}, 
in most simple terms, allows us to decompose
a bundle in terms of a ``basis''. The procedure works as follows: Start from a
sheaf $F$ on the projective space $\pp^n$, and pull it back to the product
$\pp^n\times \pp^n$. On the latter space, there is a resolution of the
structure sheaf of the diagonal, $\oo_{\Delta}$, which reads
\be
0 \to \Lambda^n \left(\oo_{\pp^n}(-1)\boxtimes \qq^* \right) \to
\ldots \to \oo_{\pp^n}(-1)\boxtimes \qq^*  \to \oo_{\pp^n\times \pp^n} 
\to \oo_{\Delta}\to 0,
\ee
where $\qq\equiv {\cal T}(-1)$ is the universal quotient bundle. We can now
tensor this resolution with $\pi_1^* F$ ($\pi_i$ will be projections on both
factors); then take an injective resolution $I^{\bullet\bullet}$ 
of this complex 
(which is, by
definition, a double complex), and apply to it the direct image of the 
second projection $\pi_{2*}$. 
Consider now the cohomology of the double
complex 
$\pi_{2*}I^{\bullet\bullet}$ 
obtained in this way; 
as usual, this can be computed by spectral sequences. There are two
spectral sequences, depending on which filtration one chooses; one gives 
the result that the cohomology of this double complex is present only in total
degree zero, and its sum is the sum of the grades of a filtration of the
original $F$; the other one has $E_1$ term
\be
E_1^{p,q}= H^p(\pp^n, F(q))\otimes \Lambda^{-q} \qq^*. \label{E1}
\ee
To be more precise: the grades here have the range $0\leq p\leq n$, $-n \leq
q\leq 0$, and the cohomology whose sum corresponds to $F$ is in grades $p=-q$.

This procedure has a clear interpretation in the framework of derived
categories \cite{Thomas1, GM} as a Fourier-Mukai transform \cite{Thomas2}. 
Indeed, the whole
process can be seen as ${\bf R}\pi_{2*}(\pi_1^*F\otimes \oo_{\Delta})$, where
${\bf R}$ is the derived functor within the derived category; and it
is clear that it is an identity from the derived category to itself.

What is interesting for us is that the double complex (\ref{E1}) has an
interpretation as complex of quivers. Let us make this more precise; 
introduce the algebra $A\equiv \Hom(
\oplus_{i=0}^n \oo_{\pp^n}(i),$ $\oplus_{i=0}^n \oo_{\pp^n}(i))$. This is the
path algebra of a quiver, that we will call {\it Beilinson quiver} (for
$\pp^n$), and which is the quiver with $n+1$ dots, $n+1$ arrows between
each pair of
consecutive dots, and relations as we described for the McKay quiver $
[X^i,X^j]=0$. Now we interpret each line of (\ref{E1}) as a representation of
this quiver as in \cite{Douglas:2000qw}, letting each $\Lambda^i \qq^*$
correspond to the $i$th irreducible representation of the quiver; 
in this way the whole double complex is an
element of the derived category $D^b(\mat{mod}-A)$ of the abelian category 
$\mat{mod}-A$ of representations of
the quiver. In fact, this construction gives 
an equivalence of
derived categories $D^b(Coh(\pp^n))$ and $D^b(\mat{mod}-A)$.

The latter can be extracted, from a complex which represents an object in the
derived category, as 
the alternated sum of its terms. 
From this we can read the
orbifold charges $n_i$, which equal $\chi(F(-i))=\chi(\oo(i),
F)\equiv(\oo(i), F)$. 

\section{Multiple fibration resolutions and tautological bundles}
\label{multiple}
As we described above, McKay correspondence uses tautological bundles
$\rrr_i$; in computations, it is in general not trivial to find them
explicitly.
A toric method to find them
from first principles was described in \cite{Diaconescu:2000ec}; it was
applied there to the \CY hypersurface $\pp^{2,2,2,1,1}[8]$. The results were very
easy as compared to the computations required to obtain them: to describe
them, let us recall \cite{Candelas:1994dm} 
that this \CY is most easily analyzed as
embedded in the resolution of the weighted projective space. This resolution
has itself a simple toric description: the charge matrix reads
\be \left[\ba{rrrrrr}
0&0&0&1&1&-2\\1&1&1&0&0&1\ea \right], \ee
which means that it is a $\pp^3$ fibration over $\pp^1$: 
$\pp\left(3\oo_{\pp^1}\oplus\oo_{\pp^1}(-2)\right)$. Let us call $H$ the
divisor corresponding to one of the first three vectors in the fan (in the
same ordering of the charge matrix), and $L$ the divisor
corresponding to the fourth (or fifth) vector. These are respectively given by
an
hyperplane in the fiber $\pp^3$, and a hyperplane in the base $\pp^1$.
In these terms, the results of \cite{Diaconescu:2000ec} 
for the $R_i\equiv \rrr_{i|\pi^{-1}(0)}$ read 
\be \ba{lcl}
R_1=\oo &\quad&R_2=\oo(L)\\
R_3=\oo(H) &\quad& R_4=\oo(H+L)\\
R_5=\oo(2H) &\quad&R_6=\oo(2H+L)\\
R_1=\oo(3H) &\quad&R_8=\oo(3H+L)
\ea\label{R1}\ee
(we changed notation with respect to 
\cite{Diaconescu:2000ec}: there the $R_i$
are ordered differently and are the duals of (\ref{R1}). This is, however,
taken into account by us by a change in the orthogonality condition). 
One immediately notices that the pattern followed is very easy: 
the coefficient of
$L$ (which is relative to $\pp^1$) goes from 0 to 1, 
and the coefficient of $H$ 
(which is relative to $\pp^3$) goes from 0 to 3.
One is naturally led to conjecture
that this relationship between $R_i$ and multiple fibration structure is
general. The precise statement 
is most easily described through an example: we will choose
$\pp^{6,2,2,1,1}[12]$, which is the companion example treated in
\cite{Candelas:1994dm}.

What we want to do is to resolve this space to obtain a multiple fibration of
projective spaces. We do that in two steps. First we resolve the locus
$z^4=z^5=0$, thus adding a new homogeneous coordinate $z^6$; then resolve
again the locus $z^2=z^3=z^6=0$. 
The final fan and charge matrices read
\be T=\left[
\ba{rrrrrrr}
1&0&0&0&-6&-3&-1\\
0&1&0&0&-2&-1&0\\
0&0&1&0&-2&-1&0\\
0&0&0&1&-1&0&0\ea\right], \quad Q=\left[ \ba{rrrrrrr}
0&0&0&1&1&-2&0\\
0&1&1&0&0&1&-3\\
1&0&0&0&0&0&1\ea\right].\label{fan}\ee
This means that the resulting space has the following structure: A fibration
in $\pp^1$ over a base, $F_{0,-2}$, which is itself a fibration (with a
notation that generalizes the
standard one for Hirzebruch surfaces) in $\pp^2$ over $\pp^1$.
In analogy with the previous case, let us call respectively $B, H$ and $L$ the
divisors corresponding to the hyperplanes in the fiber $\pp^1$, in the $\pp^2$
and in the base $\pp^1$. In analogy with what we observed in the previous
case, there should be $2\times3\times2=12$ $R_i$:
\be\ba{lcl}
R_1= \oo        & \quad &R_2=\oo(L_2)\\
R_3=\oo(L_1)    & \quad &R_4=\oo(L_1+L_2)\\
\vspace{0.2cm}R_5=\oo(2L_1)   & \quad &R_6=\oo(2L_1+L_2)\\
R_7= \oo(B)     & \quad &R_8=\oo(B+L_2)\\
R_9=\oo(B+L_1)  & \quad &R_{10}=\oo(B+L_1+L_2)\\
R_{11}=\oo(B+2L_1)& \quad &R_{12}=\oo(B+2L_1+L_2).
\ea\label{R2}\ee
This $12$ is exactly the order of the singularity we started with (let us
recall that the compact toric variety we are talking about here is the
exceptional divisor $\pi^{-1}(0)$ of the resolution of this singularity).
This is a first check of the conjecture: the fact that, for instance, the
coefficient of $H$ in (\ref{R2}) ranges from 0 to 2 is fixed, 
in the framework of
the conjecture, by the fact that it corresponds to a hyperplane in $\pp^2$.\\
It is important to note that the one we wrote down is not, obviously, the only
possible resolution of the initial weighted projective space. In particular,
it is not the same which was alluded to in 
\cite{Candelas:1994dm}; in that case there is only one step, and the
exceptional locus is a ruled surface. There are in general, indeed, other
possibilities of obtaining a multiple fibration by resolving; what makes the
resolution we chose more special, and what we believe has to do with the nice
fit $12=12$ we obtained above, is that this one does not change the canonical
class -- that is, it is {\it crepant}. We will come back to this when, at the
end, we will try to learn a lesson from the ``experimental'' discoveries we
are doing in this section.\\
We can describe, in any case, a rough motivation for the conjecture  
(apart from the suggestions
coming from (\ref{R1})). The $\rrr_i$ of the McKay correspondence are 
defined as tautological bundles on $M$, the whole non-compact resolution of
$\cc^n /\Gamma$. 
The idea is that it could even be that 
the relevant information is 
already contained 
on the exceptional locus $\pi^{-1}(0)$; if we are able to resolve
this space, in turn, in order to reveal in its interior some projective
spaces, then it is reasonable to think that the tautological bundles,
restricted to these projective spaces,
become powers of the tautological bundle on them.
We will come back again on these ideas later, after having checked the
conjecture in the example we chose, and having noted a few nice fits that make
the picture more plausible.\\

\section{Gram-Schmidt orthonormalization procedure and mutations}
\label{GS}
Before we actually do the specific computations, let us describe explicitly,
for reasons that will shortly become clear, the general Gram-Schmidt 
procedure that we
follow to find the $S_i$ such that $(R_i, S_j)= \delta_{ij}$. The first one, 
$S_1$, obviously equals $R_1$ itself. Next, let us notice that our bundles
have the property $(R_i,R_j)=0 \ \forall i>j$. 
Then the others $S_i$ can be obtained as
\bea
S_2&=& -R_2 +(R_1,R_2)R_1, \nonumber\\
S_3&=& R_3
-(R_1,R_3)R_1 -(R_2,R_3)S_2 = \\ &&R_3 -(R_2, R_3) R_2 +
\left[(R_1, R_2)(R_2,R_3)-(R_1,R_3)\right]R_1, \nonumber
\label{GS23}
\eea
and so on. Of course near the end of the series we can use Serre to make the
computations simpler.

This procedure is based only on the assumption that we made; however, it gets
a particular meaning if further properties hold for the $R_i$. Namely, let us
suppose that 
\be \Ext^k(R_i,R_j)=0 \ \ \ \forall i>j, \ \forall k \label{exc1}\ee
and that
\be \Ext^k(R_i, R_j)=0 \ \ \ \forall i\leq j, \forall k>0. \label{exc2}\ee
These conditions together
make the $R_i$, by definition, an 
{\it exceptional series}\cite{semRu}. Using the second
set of them, we can interpret $S_2$ as being given by an exact sequence
\be
0 \to S_2 \to \Hom(R_1, R_2) \otimes R_1 \to R_2 \to 0;
\label{mut}
\ee
that is, $S_2$ is the kernel of the natural evaluation. This is usually called
a {\it mutation}\cite{semRu, Zaslow:1996nk, Hori:2000ck} 
of $R_2$ within the exceptional series $\{ R_i \}$, 
more specifically a left mutation, and noted as $L R_2$.
In a similar way, we
can interpret $S_3$ as the first term in the sequence
\be
0 \to S_3 \to 
\Hom(R_1, L R_3)\otimes R_1 \to \Hom(R_2, R_3)\otimes R_2 \to R_3 \to 0;
\label{S3}
\ee
this sequence is obtained joining two sequences of the type (\ref{mut}), the
first of which is 
\be
0 \to L R_3 \to \Hom(R_2, R_3) \otimes R_2 \to R_3 \to 0
\ee
and defines $L R_3$. Using (\ref{S3}) and our assumptions we obtain 
\bea
S_3 &=& R_3 -(R_2,R_3)R_2 + (R_1, L R_3) R_1 = \nonumber\\
&&R_3 - (R_2, R_3)R_2 + \left[(R_1,R_2)(R_2,R_3) - (R_1, R_3)\right] R_1
\eea
in agreement with result for $S_3$ in (\ref{GS23}).

It is natural to ask oneself whether the series we constructed in our two
examples are exceptional. As it turns out, even more is true: they are what is
called a foundation of a {\it helix}. This means that the series $\{ R_i
\}_{i=1}^n$ can be
extended infinitely in both senses, in such a way that any $n$ consecutive
elements make up an exceptional series, and that the ``periodicity'' condition 
$R_{n+1}= R_1 \otimes K^*$ holds.

This is not only a mathematical curiosity. Being a helix is a property at the
heart of Beilinson theorem;
it is thus plausible that 
this property is crucial in this context 
to find a quiver corresponding to a given sheaf
on the \CY.
We will come back to this
later, after having checked our first conjecture for the tautological bundles
in our example $\pp^{6,2,2,1,1}$, and proved our claim that the $R_i$ are a
foundation of a helix in both examples $\pp^{2,2,2,1,1}$ and 
$\pp^{6,2,2,1,1}$.

\section{Explicit computation}
\label{expl}
\subsection{The dictionary}
We have now to check that the $R_i$ given in (\ref{R2}) really give the
correct result for the bundles on the \CY corresponding to 
the natural basis near the orbifold point. 
As a preliminary, we need some information about the divisors in this
variety and their intersections. As we said, the Picard group is generated by
three divisors $B, H, L$; we obtain the relations
\be
L^2=0, \quad H^2(H-2 L_2)=0, \quad B(B-3 H)=0, \quad B\,H^2\,L=1.
\ee
Moreover, the anticanonical divisor is $-K = 2B$. It follows that on the \CY
submanifold $Y$
\be
B_Y= 3 H_Y, \quad H^3_Y=4, \quad (H^2 L)_Y=2,
\ee
the subscript $(\ )_Y$, which we will hereafter drop when no confusion is
possible, meaning restriction to $Y$. This matches with the results of 
\cite{Candelas:1994dm}, and allows us to use results in the literature which
we will need. In particular, from now on we denote by $h$ and $l$ the
generators of the curves on $Y$, duals to $H$ and $L$ in the sense that
\be \ba{lcl}
(H\cdot h)_Y=1,&\quad& (H\cdot l)_Y=0\\
(L\cdot h)_Y=0,&\quad& (L\cdot l)_Y=1.
\ea\ee
We can now find the $S_i$, and restrict them to $Y$. 
Since we have already described the method, let us just give the final 
result for the $(S_i)_Y\equiv V_i$:
\be\ba{lcl}
\ch(V_1)=1&\quad&\ch(V_2)= -1 + L\\
\ch(V_3)=-2+H- 2 L +2 h+l+\frac23 &\quad&\ch(V_4)=2-H-l+\frac13\\
\ch(V_5)=1-H+2L+l-2h+\frac43&\quad&\ch(V_6)=-1+H-L-l-\frac13
\ea\label{Vectors}\ee
(of course the first, integer, numbers, are element of $H^0$, while the final
ones, fractionary, mean elements of $H^6$). 
We have not listed the others because $V_i+V_{6+i}=0$, a result which
parallels similar ones for the other toric varieties and that matches the
relation on the periods in the orbifold basis $\varpi_i + \varpi_{i+6}=0$.\\
From the latter relation and  
the mirror map found in \cite{Kaste:2000id,Scheidegger:2000ed}, we
find the monodromy to be 
\be\left[\ba{rrrrrr}
-1&0&1&-2&0&0\\0&1&0&0&2&0\\-1&1&-1&-1&2&1\\1&0&0&1&0&0\\-1&0&0&-1&1&1\\
1&0&0&1&0&-1\ea\right]\ee
and as a consequence, acting repeatedly on the pure pure D6-brane state, we
obtain the states
\be\ba{lcc}
v_1=(1,0,0,0,0,0)&&\\
v_2=(-1,0,1,-2,0,0)&&\\
v_3=(-2,1,-2,-1,2,1)&&\\
v_4=(2,-1,0,4,0,-1)&&\\
v_5=(1,-1,2,-1,-2,1)&&\\
v_6=(-1,1,-1,-2,0,-1)&&
\ea\label{vectors}\ee
and their negatives. The charges in (\ref{vectors}) are listed as $(n_6,n_4^1,
n_4^2,n_0,n_2^1,n_2^2)$; to complete the check, we have to compare these
charges with the ones in the Chern polynomials. This is accomplished as usual
by comparing the central charges in the two bases: $Z=n\cdot\Pi=-\int e^{-t}
\ch(V)\sqrt{{\hat A}(T)/{\hat A}(N)}$, where $t$ is the complexified
K\"ahler form. We find
\be
r= n_6,\quad c_1= n_4^1 H+ n_4^2 L,\quad \ch_2= n_2^1 h+ n_2^2 l ,\quad 
-\ch_3=n_0+\frac{13}2n_4^1+2n_4^2,
\ee
using which the check can now be easily completed, comparing (\ref{Vectors}) 
and (\ref{vectors}).

\subsection{Helices}
We will describe now the proof of the claims given at the end of section
\ref{multiple}, that the $R_i$ make up a foundation of a helix both in the
example of this paper and in the one given in \cite{Diaconescu:2000ec}. 
We will limit ourselves to describe the main ideas, skipping details when they
become too technicals.

The latter case is easier, so let us start by that one. What we have to do is
to compute cohomology groups in toric geometry; there is a standard method to
do that \cite{Fulton}, 
but we find it easier (and perhaps more instructive) to use
a mix of this and of other techniques. In the terminology of
\cite{Diaconescu:2000ec}, we have to check that the bundles $k H + L, k H$
for $k=0,\ldots,3$ and $k H-L$ for $k=1,2,3$ enjoy the property (which we will
call acyclicity) $h^i=0 \ \forall i >0$. \\
First of all, we will use a consequence of the general method \cite{Fulton}: a
sheaf generated by its sections on a toric variety is acyclic. The condition
for this to be true, by the general theory, turns out to include all bundles
$kH +L$ and $kH$, but not $kH -L$. To treat these, we use an exact sequence 
\bea
0 \to \oo\left((k-1)H +L \right) &\to&
\oo(kH-L)\to \\
&& \oo_{|(H-2L)}(kH-L) \cong
\oo_{\pp^1}(k)\boxtimes\oo_{\pp^2}(-1)\to 0;\nonumber\label{funseq}
\eea
since $\oo_{\pp^2}(-1)$ has no cohomology, we reduce to the case previously
treated. \\
We have to prove, in addition, that the inverses of these line bundles have no
cohomology. We use this time two kinds of exact sequences, the first kind in
which again the restriction to the divisor $H-2L$ appears, the second one in
which instead the restriction is to $L$. Using these, we can by a zig-zag
procedure prove the result for all the bundles we need; a relevant feature 
is that one sees that there is a range of negative bundles with this
property that is just enough bug to include those we are interested in. This
is analogous to what happens with the series $\oo(1), \ldots \oo(n)$ in
$\pp^n$: in that case the negative bundles from $\oo(-1)$ to $\oo(-n)$ have
no cohomology, and $\oo(-n-1)$ starts to have it. It seems as if the
conjectural method we described to find the $R_i$ can be viewed as a means to
construct helices on multiple fibrations of a certain type. We will see later
why this is non trivial.

The second case, $\pp^{6,2,2,1,1}$, is more complicated, but is conceptually
similar, and we will be very sketchy. 
The bundles that we have to prove to be acyclic are now: a) $k H + L, k H$
for $k=0,1,2$ and $k H-L$ for $k=1,2$; b) $B+kH+ k^\prime L$, 
for $k=-2,\ldots,2$
and $k^\prime=-1,0,1$. The case a) can 
be reduced to an analysis on a reduced fan
which is nothing but a 3-dimensional analog of the $\pp^{2,2,2,1,1}$ case that
we have just seen. The case b) makes use again of a sequence very similar to 
(\ref{funseq}), with the restriction to the divisor $B$ appearing instead of
that to $H-2L$. In that case, the divisor turned out to be isomorphic to
$\pp^1\times \pp^2$, and we had at our disposal known vanishing theorems
for line bundles on $\pp^n$; in this case, the divisor $B$ is isomorphic to
$\pp^{2,2,1,1}$, and again we can use a 3-dimensional analog of the discussion
above to get the vanishing theorems needed. \\
Finally, we have to show that the bundles inverse to those of the cases a) and
b) have no cohomology. Similar techniques to the above let us get the
desired results for these bundles as well.

In both cases, what we have really shown is that the $R_i$ are an exceptional
collection. To show that they are a foundation of a helix requires to check
that $R^{(n-1)}R_1=R_1(-K)$, where $n$ is the length of the collection. About
the proof of this we have nothing special to say, but that it is made easier
by reformulating in terms of both left and right mutations; and that it is
this last fact is very plausible from the very beginning, due to the
peculiarly easy form of the series. In the $\pp^{2,2,2,1,1}$ case, indeed, let
us observe that, if one were to guess the term following $R_8$ in the series
(\ref{R1}), one would naturally write $\oo(4H)$, which is indeed $K^*$; the
same is true for the case of $\pp^{6,2,2,1,1}$, for which the natural guess
after the series (\ref{R2})would be $\oo(2B)$, which is again $K^*$.

\section{McKay and Beilinson}
\label{mkb}
Let us finally
analyze the relationship
between the picture that emerged in this paper and the McKay correspondence.
The McKay correspondence, as we saw, gives the $S_i$ as complexes in terms of
the tautological bundles $\rrr_j$ on the resolution of the singularity.
The method emerged in this paper gives them, instead, intrinsically in terms
of the exceptional locus, exploiting the existence of 
a helix on this locus to use the technique of mutations. The fact that both
methods work is in itself a proof of the fact that they are compatible;
however, a systematic comparison could be interesting to do.

To give an idea of what this comparison looks like, we treat the easy case of
projective spaces $\pp^n$. In this case,  
 the McKay quiver has $n+1$ dots
arranged cyclically
with $n$ links between each pair of
consecutive dots (to form a closed loop), and with the usual relations;
Beilinson one, as we said,
is the same but with links between dot $\{n+1\}$ and $\{1\}$ 
missing. 
Moreover, $R_i=\rrr_{i|\pp^n}=\oo_{\pp^n}(i-1)$
. Throughout this example, we
will keep a notation halfway from explicit and abstract, exploiting only the
form of the quiver, to emphasize how to do the comparison in general.
In the McKay framework,
\be
S_1 = \rrr_1 - 
a^{(1)}_{21} \rrr_2 + a^{(2)}_{31} \rrr_3 + \ldots +(-)^n \rrr_1
\label{S1MK}
\ee
as a K-theory class. We know \cite{INa} that this class has support only on
$\pp^n$.
From the definition of helices, on the other hand, we know that the $n$th
right mutation of a helix gives the helix itself. This gives in general the 
complex
\bea
0\to R_i &\to &\Hom(R_i, R_{i+1})^* \otimes R_{i+1}\to
 \Hom(R^{(2)}R_i, R_{i+2})^* \otimes R_{i+2}\to \nonumber\\
\ldots&\to & \Hom(R^{(n-1)}R_i, R_{i+n})^* \otimes R_{i+n}\to R_i(-K)\to 0
\label{relation}
\eea
(which in this case is simply $\oo - V \oo(1) + \Lambda^2 V \oo(2)+\ldots
+(-)^{n+1}\oo(n+1)=0$, where $V=H^0(\pp^n, \oo(1))$). Remember now that the
noncompact manifold $M$, which is the resolution of the singularity, is the
total space of the line bundle $K$ on the exceptional locus.
Then we can pull back
(\ref{relation}) to $M$, to find an analogous 
relation on the $\rrr_i$. Exploiting the fact that 
$a_{21}^{(1)}=V^*=\Hom(R_1,R_2)^*$, 
$a_{i+1,1}^{(i)}=\Lambda^i V^*=\Hom(R^{(i-1)}R_1,R_{i+1})^*$
we can use the pull-back of (\ref{relation}) for $i=1$ 
on $M$ to
reexpress first $n$ terms (that is, all outside the last) of (\ref{S1MK}) as
$S_1= \rrr_1 - \rrr_1(-K)$.\\
The pull-back $p^*K$ of $K$ 
to its total space $M$ has a tautological
section which has a zero exactly on $\pp^n\subset M$. This gives a sequence
\be
0 \to p^* K^*  \to \oo_M \to \oo_{\pp^n} \to 0;
\ee
by twisting it with $\rrr_1$, we obtain
\be
S_1= \rrr_1 - \rrr_1(-K)= \rrr_{1|\pp^n}=R_1;
\ee
that is, we obtain that the result is supported on $\pp^n$, as it should, and 
the same result as with the mutation method (although for this first step it
is a little trivial).

A less trivial check is obtained with the second term; one uses again the
pull-back of (\ref{relation}), this time with $i=2$. Exploiting 
$a_{12}^{(1)}=\Lambda^{n-1}V^*=V=\Hom(R_1,R_2)$, 
$a_{i+2,2}^{(i)}=\Lambda^i V^*=\Hom(R^{(i-1)}R_2,R_{i+2})^*$, we can 
reexpress this time
first $n-1$ terms of 
\be
S_2 = \rrr_2 - 
a^{(1)}_{32} \rrr_3 + a^{(2)}_{42} \rrr_4 + \ldots +(-)^n \rrr_2
\label{S2MK}
\ee
as 
\be 
S_2 = \rrr_2 - V \rrr_1 - (\rrr_2(-K) - V\rrr_1(-K)) = R_2 - (R_1,R_2) R_1
\ee
as it should be. It is straightforward to continue to the end this check.
Let us note, by the way, 
that a similar method can give compact expressions for 
the $V_i$ on the \CY in terms of the $R_i$.

Thus, we have shown in this example that Beilinson and McKay methods agree and
give the same $S_i$; in doing that, we have exploited a series or relations
between $a^{(i)}$ and $\Hom(R_j, R_k)$ that we can resume by 
saying that Beilinson quiver is
invariant under mutations, and that McKay quiver is an extension of it, in the
sense that Beilinson one is obtained by cutting arrows between two consecutive
dots: this is the mathematical counterpart of what was done in
\cite{Diaconescu:2000ec}. In this case we already knew this fact (it was 
stated at the beginning of this section), but it is true in general; this is
essential, as it allows to interpret the quiver representation 
given by Bondal construction
we outlined in section \ref{disc} as a representation of the McKay quiver, and
hence as a 
D-brane. Moreover, this fact seems to support and generalize the claim
\cite{Douglas:2000qw}
that all holomorphic
objects near the orbifold points are given by Beilinson representations,
i.e. by representations of the McKay quiver whose links 
from the last dot and to the first one have been cut. 

As a further clarification of this phenomenon, let us describe an example,
again for the $\pp^n$ case.  
The
representation of the McKay quiver with all $n_i=1$ is the D0 on the
resolution $M$. On the other hand, one can easily see from its very definition
as an universal quotient, that, chosen a point $p \in \pp^n$, 
the bundle $\qq$ has a section which vanishes exactly in $p$. From this we
obtain a resolution 
\be
0 \to \Lambda^n \qq^* \to \ldots \to \qq^* \to \oo_{\pp^n} \to \oo_p\to 0,
\ee
which gives $\oo_p = \oplus S_i = \oplus n_i S_i$, with
$n_i=1$. The representation of the Beilinson quiver corresponding to the McKay
one describes the same state but bound not to move from the exceptional locus.
In this way one can, by the way, prove the relations mentioned shortly after
(\ref{Vectors}).

\section{Discussion}
\label{disc}
Let us now try to put all the pieces together. What we initially tried was a
guess for the K-theory classes corresponding, in the large volume limit, to
the orbifold basis of D-branes. In doing that, we noticed that the $R_i$
are not only so easy to guess in general without doing the complicated
computations starting from first principles as in \cite{Diaconescu:2000ec},
but that they are the foundation of a helix. This cannot be a coincidence, and
we want now to justify this fact a posteriori, or at least to see it from a
larger perspective.

First of all, let us expand the comment about helices and Beilinson theorem
that we made at the end of section \ref{GS}. We saw in section \ref{Bth}
that, on projective
spaces, Beilinson procedure takes an element of the 
derived category and expresses it as the
cohomology of a double complex, which we may then use to find a quiver, and
thus the ``translation'' of the initial sheaf in orbifold terms. There were,
in that proof, two relevant series of bundles: the $\oo(i)$ and the
$\Lambda^j \qq^*$. 
Note that these are precisely the $R_i$ and 
$(-)^j S_j$ in the case
of the projective space. Of course, as far as general definitions of these two
series of objects are concerned, we only know that the first ones are
restrictions to $\pi^{-1}(0)$ of the tautological bundles, and the second ones
are dual to them; we do not know a priori of any reason for which they could
give a resolution. But, this reason is just what we found, almost by chance
(see discussion below), to be true in our cases: it is the extra piece of
information {\sl that $R_i$ are foundation of a helix} 
(a first, though partial, proof of
this fact is a theorem \cite{GoRu}, which still refers to $\pp^n$, but whose
proof already seems independent enough from this assumption; a more general, 
though more abstract, construction, is then given in the paper by Bondal in
\cite{semRu}). More explicitly, 
the definition of Beilinson quiver 
we gave for $\pp^n$ can be generalized as the quiver whose path algebra is 
$A \equiv \Hom(\oplus_i R_i, \oplus_i R_i)$. Then, if the $R_i$ are 
a foundation of
a helix, there is an equivalence of 
derived categories $D^b(Coh(\pi^{-1}(0))) \cong D^b(\mat{mod}-A)$, as for the
case in which $\pi^{-1}(0)=\pp^n$ that we saw above. This is the
generalization of Beilinson theorem we needed, and it needs the helix property
that we found!

Let us now come back to our multiple fibration resolution. We
already noted that the match between the number of $R_i$ found by our method
and the order of $\Gamma$ is already a non trivial check. We add now that
another non trivial fact is that we have found a helix on the multiple
fibration. The simplest examples of multiple fibrations are given by
Hirzebruch surfaces $F_k$; in this case \cite{semRu}, it is known that, in the
cases $F_k, k>3$ there are no helices made up of line bundles. So our examples
could seem to be special in two senses: 1) they yield the match we talked
about above; 2) they allow helices on them made up of line bundles. 
One special property that our resolutions share is that they are
crepant resolutions of the original $\pi^{-1}(0)$. This is linked with 
another, very natural, property that toric varieties can have: that of 
allowing a non-singular \CY inside them. In general, indeed, it is true
that one can take in any toric variety a subvariety supported on the
anticanonical divisor, and that a so chosen subvariety 
has formally a trivial canonical bundle; but, for most ambient toric
varieties, there would be no way to find a non singular \CY in this way.
The condition to find non singular \CY is that the
polyhedron of the toric variety, with respect to its anticanonical sheaf,
be integral; and this condition in turn means that the ambient toric variety
has only Gorenstein singularities, which admit a partial crepant 
resolution \cite{bat}. 

So we have found that an event which seems a priori to be very 
unlikely, the existence on a multiple toric fibration 
of a helix with the right properties, seems to 
take place precisely when the toric fibration admits a 
non singular \CY inside. This fact
is somewhat surprising, from a mathematical point of view;
the reason is probably that the dictionary between
sheaves on the large volume \CY and quivers is required to exist by physics.
Although we concentrated in this paper on a class of examples (in general, as
stated above, the crepant resolution is only partial; and the final space 
after the resolution could
be different from a multiple fibration), the structure found here makes it 
probable that helices play a role in more general cases as well 
\cite{mayr,ind}.

A final remark is that we could have
even guessed that our dictionary makes use of helices, because of their
mirror symmetry interpretation \cite{Hori:2000ck, Hori:2000kt}. Let us  
check whether we obtain a consistent result even from this point of view.

\subsection{The mirrors of the helices}

The mirror theory to the linear sigma model we are dealing with was found in 
\cite{Hori:2000kt} by a dualization procedure inspired by T-duality from the
world-sheet perspective. Let us put aside, in this
section, the superpotential. Before dualizing, let us modify the initial 
linear sigma
model to take in account the resolutions of the exceptional locus. The initial
model has a $U(1)$ gauge invariance and 5 chiral multiplets $Z^i$ (apart from
$P$); this describes, as the exceptional locus, the toric variety $\pp^{w_1,
  \ldots, w_5}$. Since we resolved the latter
 in a more general toric variety, we
modify the charge matrix of the model in exactly the same way in which we 
modified the charge matrix of the toric manifold, adding 
chiral multiplets and gauge invariances. For instance, in the
$\pp^{6,2,2,1,1}$ case, we add two chiral and two gauge multiplets. 

The result we need is that the dual theory
has a twisted superpotential of the form 
\be
\widetilde W = \sum_{a}\Sigma_a \left( \sum_i Q_{ai} Y^i - \rho_i \right) + 
\mu \sum_i e^{-Y^i}
\ee
where $\Sigma_a$ are the gauge multiplets, $Y^i$ are the dual twisted chiral
multiplets, and $\rho_i$ are the FI terms. In the $\rho_i\to \infty $ limit,
which in the original theory is the sigma model limit, the gauge multiplets
become infinitely heavy and become Lagrange multiplier: we end up with a
twisted superpotential $\widetilde W = \mu \sum_i e^{-Y^i}$ and constraints 
$\sum_i Q_{ai} Y^i = \rho_i$. In this dual theory, branes are described
\cite{Hori:2000ck} by their images in the complex plane under the map defined
by the superpotential: each brane gets mapped to a straight half-line coming
out from a critical point. A physical analysis of branes in this theory allows
then to see the counterparts of the helix condition -- although a strange
feature seems to be that one has to break supersymmetry at some stage.\\
The simple check we want to do here is that there
is the right number of critical points - and thus, the right number of mirror
branes. We will do that again for the $\pp^{6,2,2,1,1}$ example, trying to
describe the features that give the expected agreement.
Defining $\lambda_i=e^{-\rho^i}$, and Lagrange multipliers $\alpha_i$,
we obtain from the charge matrix (\ref{fan}) a system of equations:
\be
\lambda_3=\alpha_3 (\alpha_3 -3 \alpha_2), \quad \lambda_2=
\frac{\alpha_2^2(\alpha_2 -2 \alpha_1)}{ (\alpha_3 -3 \alpha_2)^3}, \quad
\lambda_1  = \frac{\alpha^2_1}{(\alpha_2 - 2 \alpha_1)^2}.
\label{lagrange}
\ee
This system has degree $2\times3\times2=12$; the critical values of the
superpotential equal 
$\widetilde W = 2 \alpha_3$, which has thus $12$ solutions, as
it should. Explicitly we have $\alpha_3= 
\sqrt{ \lambda_3\left( 1 + 3\sqrt[3]{\lambda_2 (1+2\sqrt{\lambda_1}) } \right)}$, where each of the roots is understood with
multiple choice. So the 12 critical points are organized generically in 6
circles in the complex $\widetilde W$ plane. \\
This agreement is due to the following fact.
The degrees of the three equations are dictated by the three rows of
the charge matrix in (\ref{fan}). If the negative number in the second row,
for instance, were different from $-3$, the exponent of the denominator in the
second equation in (\ref{lagrange}) would be higher than 3, and the
resulting would be higher, giving more critical points than
needed. The
$2\times3\times2$ which came in our conjecture because of the dimensions of
the $\pp^k$ in the multiple fibration structure, here is transformed in the
same product, but with each factor coming from the degree of an equation.
These are equal because all rows of the charge matrix but the one
corresponding to the fiber have zero sum of degrees, as one can check in
examples. This seems to be connected again 
with the integrality of the polyhedron we referred to in the above discussion.

\smallskip {\bf Note added.} When this paper was ready for
publication, a paper
appeared \cite{ind} which overlaps with this work.

\end{document}